\documentclass[12pt]{article}

\usepackage{amsmath}

\DeclareMathAlphabet\mathbb  {U}{msb}{m}{n}
\DeclareFontFamily{U}{msb}{} \DeclareFontShape{U}{msb}{m}{n}{
  <5> <6> <7> <8> <9> gen * msbm
  <10> <10.95> <12> <14.4> <17.28> <20.74> <24.88> msbm10
  }{}

\renewcommand{\title}[1]{\null\vspace{10mm}\noindent
                         {\Large{\bf #1}}\vspace{10mm}}
\newcommand{\authors}[1]{\noindent{\large #1}\vspace{20mm}}
\newcommand{\address}[1]{\center{\noindent\small\itshape #1\vspace{0mm}}}

\makeatletter
\def\section{\@startsection{section}{1}{\z@}{-3.25ex plus -1ex minus
             -.2ex}{1.5ex plus .2ex}{\normalfont\bfseries}}
\def\subsection{\@startsection{subsection}{1}{\z@}{-3.25ex plus -1ex
                minus -.2ex}{1.5ex plus .2ex}{\normalfont\itshape}}

\renewenvironment{thebibliography}[1]
         {\section*{References}\frenchspacing\small
          \begin{list}{[\arabic{enumi}]}
         {\usecounter{enumi}\parsep=2pt\topsep 0pt
         \settowidth{\labelwidth}{[#1]}
         \leftmargin=\labelwidth\advance\leftmargin\labelsep
         \rightmargin=0pt\itemsep=0pt\sloppy}}{\end{list}}

\makeatother

\begin{document}

\thispagestyle{empty}

\begin{titlepage}

\begin{center}
\hspace*{\fill}{{\normalsize \begin{tabular}{l}
                              \textsf{hep-th/0110231}\\
                              \textsf{REF. TUW 01-025}\\
                              \textsf{REF. UWThPh-2001-40}
                              \end{tabular}   }}

\title{Noncommutative spin-$\frac{1}{2}$ representations}

\authors {{  J.~M.~Grimstrup$^{1}$, 
H.~Grosse$^{2}$, E.~Kraus$^{3}$,\\[1mm] 
L.~Popp$^{4}$, M.~Schweda$^{5}$, R.~Wulkenhaar$^{6}$ }}  

\vspace{-10mm}
       
\address{$^{1,4,5}$  Institut f\"ur Theoretische Physik, 
Technische Universit\"at Wien \\
Wiedner Hauptstra\ss e 8--10, A-1040 Wien, Austria}
\address{$^{2,6}$  Institut f\"ur Theoretische Physik, 
Universit\"at Wien\\
Boltzmanngasse 5, A-1090 Wien, Austria}
\address{$^{3}$ Physikalisches Institut der Universit\"at Bonn,\\
          Nu\ss{}allee 12, D-53115 Bonn, Germany }

\footnotetext[1]{jesper@hep.itp.tuwien.ac.at, work supported by 
The Danish Research Agency.}

\footnotetext[2]{grosse@doppler.thp.univie.ac.at.}

\footnotetext[3]{kraus@th.physik.uni-bonn.de.}

\footnotetext[4]{popp@hep.itp.tuwien.ac.at, work supported in part 
by ``Fonds zur F\"orderung der Wissenschaftlichen Forschung'' (FWF) 
under contract P13125-PHY.}

\footnotetext[5]{mschweda@tph.tuwien.ac.at.}

\footnotetext[6]{raimar@doppler.thp.univie.ac.at, Marie-Curie Fellow.}

\vspace{15mm}

\begin{minipage}{12cm}
  
  {\it Abstract.} In this letter we apply the methods of our previous
  paper \texttt{hep-th/0108045} to noncommutative fermions. We show
  that the fermions form a spin-$\frac{1}{2}$ representation of the
  Lorentz algebra.  The covariant splitting of the conformal
  transformations into a field-dependent part and a $\theta$-part
  implies the Seiberg-Witten differential equations for the fermions.

\vspace*{1cm}
\end{minipage}

\end{center}

\end{titlepage}

\section{Introduction}

This letter is an extension of ideas of our previous paper
\cite{Bichl:2001yf} to noncommutative fermion fields. We define a
noncommutative version of (infinitesimal, rigid) conformal
transformations and show that they leave the noncommutative Dirac
action invariant.  The conformal operators and noncommutative gauge
transformations form a Lie algebra (a semidirect product). However,
since gauge transformations involve the $\star$-product (and thereby
the noncommutative parameter $\theta$) one immediately sees that the
individual conformal operators (field- and $\theta$-transformations)
commuted with gauge transformations do not close in the above Lie
algebra. There exists a certain splitting of the combined conformal
operators into new individual components so that the commutator of
gauge transformations with them is again a gauge transformation.
{}From this new splitting we derive the Seiberg-Witten differential
equations for the fermion fields.

Finally a comment on Lorentz transformations. In general one should
distinguish between {\it observer} and {\it particle} Lorentz (or more
general, conformal) transformations, which are inequivalent when one
considers background fields equipped with Lorentz indices. In the
following we solely refer to `observer' Lorentz transformations.
Please refer to \cite{Bichl:2001yf} and references therein for further
details.

\section{Commutative case} \label{comm}

We recall from our previous paper \cite{Bichl:2001yf} the commutative 
Ward identity operators of
primitive conformal\footnote{In the following the term `conformal'
will always refer to the rigid 
transformations as introduced in \cite{Bichl:2001yf}.}
 transformations of the gauge field $A_\mu$:
\begin{align}
W^T_{A;\tau}  &:= \int d^4x\,\mathrm{tr} 
\Big(\partial_\tau A_\mu \,\frac{\delta}{\delta A_\mu}\Big) ,
  \label{TA}
\\
W^R_{A;\alpha\beta} &:=
\int d^4x\,\mathrm{tr} \Big(
\big(g_{\mu\alpha} A_\beta -g_{\mu\beta} A_\alpha 
+ x_\alpha \partial_\beta A_\mu - x_\beta \partial_\alpha A_\mu \big)
\,\frac{\delta}{\delta A_\mu}\Big),
\label{RA}
\\
W^D_A &:= \int d^4x\,\mathrm{tr} \Big(
\big( A_\mu + x^\delta \partial_\delta A_\mu\big)\,
\frac{\delta}{\delta A_\mu}\Big).
\label{DA}
\end{align}
The commutative (primitive) conformal transformations of fermions $\psi$ and $\bar{\psi}=\psi^\dag \gamma^0$ are
given by
\begin{align}
W^T_{\psi;\tau} &= \int d^4x \, \bigg( \Big\langle 
\frac{\overleftarrow{\delta}}{\delta \psi} \,\partial_\tau \psi \Big\rangle 
+ \Big\langle 
\partial_\tau \bar{\psi} \,\frac{\overrightarrow{\delta}}{
\delta \bar{\psi}} \Big\rangle \bigg) ~, 
\label{WTF}
\\
W^R_{\psi;\alpha\beta} &= \int d^4x \, \bigg( 
\Big\langle \frac{\overleftarrow{\delta}}{\delta \psi} \Big(
x_\alpha \partial_\beta \psi 
- x_\beta \partial_\alpha \psi 
+ \frac{1}{4} [\gamma_\alpha,\gamma_\beta] \psi \Big)  \Big\rangle
\nonumber
\\
& \hspace*{5em} 
+ \Big\langle \Big( x_\alpha \partial_\beta \bar{\psi}
- x_\beta \partial_\alpha \bar{\psi}
- \frac{1}{4} \bar{\psi} [\gamma_\alpha,\gamma_\beta] \Big)
\frac{\overrightarrow{\delta}}{\delta \bar{\psi}}  \Big\rangle
\bigg)~, 
\label{WRF}
\\
W^D_{\psi}  &=  \int d^4x \, \bigg( 
\Big\langle \frac{\overleftarrow{\delta}}{\delta \psi} \Big(
\frac{3}{2} \psi + x^\delta \partial_\delta \psi  \Big)  \Big\rangle
+
\Big\langle \Big(\frac{3}{2} \bar{\psi} 
+ x^\delta \partial_\delta \bar{\psi}  \Big)
\frac{\overrightarrow{\delta}}{\delta \bar{\psi}}  \Big\rangle
\bigg)~,
\label{WDF}
\end{align}
where the bracket $\langle\,,\,\rangle$ indicates the invariant
product in spinor space and the trace in colour space.

The Dirac action
\begin{align}
  \Sigma_D = \int d^4x\, \big\langle \bar{\psi} 
 \big(\mathrm{i}  \gamma^\mu(\partial_\mu - \mathrm{i}A_\mu) -m\big) 
\psi \big\rangle~,
\end{align}
is invariant under a gauge transformation $W^G_{A+\psi;\lambda}$ with 
\begin{align}
W^G_{A+\psi;\lambda} &= W^G_{A;\lambda} +W^G_{\psi;\lambda} ~,
\nonumber
\\
W^G_{A;\lambda} &= \int d^4x \,\mathrm{tr} 
\Big( \big(\partial_\mu \lambda - \mathrm{i}[A_\mu,\lambda]\big)
\,\frac{\delta}{\delta A_\mu} \Big) ~,
\nonumber
\\
W^G_{\psi;\lambda} &= \int d^4x \Big(
\Big\langle (-\mathrm{i} \bar{\psi} \lambda )\,
\frac{\overrightarrow{\delta}}{\delta \bar{\psi}} \Big\rangle 
+
\Big\langle 
\frac{\overleftarrow{\delta}}{\delta \psi} \,
(\mathrm{i} \lambda \psi) \Big\rangle \Big)~.
\label{gauge}
\end{align}
Furthermore, we have invariance under translation $W^T_{A+\psi;\tau}$ and 
rotation 
$W^R_{A+\psi;\alpha\beta}$ and additionally, 
for $m=0$, under dilatation $W^D_{A+\psi}$, where
\begin{align}
W^T_{A+\psi;\tau} &= W^T_{A;\tau} + W^T_{\psi;\tau} ~, &
W^R_{A+\psi;\alpha\beta} &= W^R_{A;\alpha\beta} 
+ W^R_{\psi;\alpha\beta} ~, &
W^D_{A+\psi} &= W^D_{A} + W^D_{\psi} ~.
\end{align}

\section{Noncommutative case} \label{noncomm}
\label{sec6}

The noncommutative generalization is obvious. The noncommutative
conformal transformations of the gauge field \cite{Bichl:2001yf} and
fermions are given by
\begin{align}
W_{\hat{A};\tau}^T &:= 
\int d^4x\,\mathrm{tr}\Big( \partial_\tau \hat{A}_\mu \,
\frac{\delta}{\delta \hat{A}_\mu} \Big) ~,
\label{WTTA}
\\
W_{\hat{A};\alpha\beta}^R &:= 
\int d^4x\,\mathrm{tr}\Big( 
\Big(\frac{1}{2} \big\{ x_\alpha, 
\partial_\beta \hat{A}_\mu \big\}_\star 
- \frac{1}{2} \big\{ x_\beta, 
\partial_\alpha \hat{A}_\mu \big\}_\star 
+ g_{\mu\alpha} \hat{A}_\beta
- g_{\mu\beta} \hat{A}_\alpha \Big)
\,\frac{\delta}{\delta \hat{A}_\mu} \Big) ~,
\label{WTRA}
\\
W_{\hat{A}}^D &:= 
\int d^4x\,\mathrm{tr}\Big( 
\Big(\frac{1}{2} \big\{ x^\delta, 
\partial_\delta \hat{A}_\mu \big\}_\star 
+ \hat{A}_\mu \Big) \,\frac{\delta}{\delta \hat{A}_\mu} \Big) ~,
\label{WTDA}
\\[1ex]
W^T_{\hat{\psi};\tau} &:= \int d^4x \,  \bigg( 
\Big\langle \frac{\overleftarrow{\delta}}{\delta \hat{\psi}} \,
\partial_\tau \hat{\psi} \Big\rangle
+ \Big\langle 
\partial_\tau \Hat{\Bar{\psi}} \,\frac{\overrightarrow{\delta}}{
\delta \Hat{\Bar{\psi}}} \Big\rangle \bigg) ~, 
\label{WHTF}
\\
W^R_{\hat{\psi};\alpha\beta} &:= \int d^4x \,  \bigg( 
\Big\langle \frac{\overleftarrow{\delta}}{\delta \hat{\psi}} \Big(
x_\alpha \star \partial_\beta \hat{\psi} 
- x_\beta \star \partial_\alpha \hat{\psi} 
- \frac{\mathrm{i}}{2} \theta_\alpha^{~\rho} \partial_\rho \partial_\beta
\hat{\psi} 
+ \frac{\mathrm{i}}{2} \theta_\beta^{~\rho} \partial_\rho \partial_\alpha
\hat{\psi} 
+ \frac{1}{4} [\gamma_\alpha,\gamma_\beta] \hat{\psi} \Big) \Big\rangle
\nonumber
\\
& \hspace*{4em} 
+ \Big\langle 
\Big( \partial_\beta \Hat{\Bar{\psi}} \star x_\alpha
-\partial_\alpha \Hat{\Bar{\psi}} \star x_\beta
+ \frac{\mathrm{i}}{2} \theta_\alpha^{~\rho} \partial_\rho \partial_\beta
\Hat{\Bar{\psi}} 
- \frac{\mathrm{i}}{2} \theta_\beta^{~\rho} \partial_\rho \partial_\alpha
\Hat{\Bar{\psi}} 
- \frac{1}{4} \Hat{\Bar{\psi}} [\gamma_\alpha,\gamma_\beta] \Big)
\frac{\overrightarrow{\delta}}{\delta \Hat{\Bar{\psi}}}
\Big\rangle
\bigg)~, 
\label{WHRF}
\\
W^D_{\psi}  &:=  \int d^4x \, \bigg( 
\Big\langle 
\frac{\overleftarrow{\delta}}{\delta \hat{\psi}} \Big(
\frac{3}{2} \hat{\psi} + x^\delta \star 
\partial_\delta \hat{\psi} 
\Big) \Big\rangle
+
\Big\langle
\Big(\frac{3}{2} \Hat{\Bar{\psi}} + 
\partial_\delta \Hat{\Bar{\psi}} \star x^\delta  \Big)
\frac{\overrightarrow{\delta}}{\delta \Hat{\Bar{\psi}}}
\Big\rangle
\bigg)~.
\label{WHDF}
\end{align}
The noncommutative Dirac action is defined by
\begin{align}
\hat{\Sigma}_D = \int d^4x\, \big\langle \Hat{\Bar{\psi}} 
\star \big(\mathrm{i}  \gamma^\mu \hat{\mathrm{D}}_\mu -m\big) 
\hat{\psi} \big\rangle~,
\label{NCD}
\end{align}
where 
\begin{align}
\hat{\mathrm{D}}_\mu \hat{\psi} &= \partial_\mu \hat{\psi} -
  \mathrm{i} \hat{A}_\mu \star \hat{\psi}~, &
\Hat{\Bar{\mathrm{D}}}_\mu \Hat{\Bar{\psi}} &= 
\partial_\mu \Hat{\Bar{\psi}} + 
  \mathrm{i} \Hat{\Bar{\psi}} \star  \hat{A}_\mu 
\end{align}
are the noncommutative covariant derivatives in the (anti)fundamental 
representation. The action
(\ref{NCD}) is invariant under noncommutative gauge transformations 
\begin{align}
W^G_{\hat{A}+\hat{\psi};\hat{\lambda}} &= 
W^G_{\hat{A};\hat{\lambda}} +W^G_{\hat{\psi};\hat{\lambda}} ~,
\nonumber
\\
W^G_{\hat{A};\hat{\lambda}} &= \int d^4x \,\mathrm{tr} 
\Big( \hat{D}_\mu \hat{\lambda} 
\,\frac{\delta}{\delta \hat{A}_\mu} \Big) ~,
\nonumber
\\
W^G_{\hat{\psi};\hat{\lambda}} &= \int d^4x \bigg(
\Big\langle (-\mathrm{i} \Hat{\Bar{\psi}} \star \hat{\lambda} )\,
\frac{\overrightarrow{\delta}}{\delta \Hat{\Bar{\psi}}} \Big\rangle 
+
\Big\langle 
\frac{\overleftarrow{\delta}}{\delta \hat{\psi}} \,
(\mathrm{i} \hat{\lambda} \star \hat{\psi}) \Big\rangle \bigg)~,
\label{nc-gauge}
\end{align}
where $\hat{D}_\mu \bullet = \partial_\mu \bullet - \mathrm{i}
[\hat{A}_\mu, \bullet]_\star$ is the noncommutative covariant
derivative in the adjoint representation. 

Let us first compute the rotational transformation of the action
(\ref{NCD}). We find:
\begin{align*}
(W^R_{\hat{A};\alpha\beta} + W^R_{\hat{\psi};\alpha\beta})
\hat{\Sigma}_D 
& = \int d^4x \, \bigg( \theta_\alpha^{~\rho} \Big(
- \frac{\mathrm{i}}{2} \big\langle \Hat{\Bar{\psi}} \star 
\gamma^\mu \partial_\rho \hat{A}_\mu \star 
\partial_\beta \hat{\psi} \big\rangle
+ \frac{\mathrm{i}}{2} 
\big\langle \Hat{\Bar{\psi}} \star \gamma^\mu \partial_\beta
\hat{A}_\mu \star \partial_\rho \hat{\psi} \big\rangle
\Big)
\\
& \hspace*{3em} + \theta_\beta^{~\rho} \Big(
 \frac{\mathrm{i}}{2} \big\langle \Hat{\Bar{\psi}} \star 
\gamma^\mu \partial_\rho \hat{A}_\mu \star 
\partial_\alpha \hat{\psi} \big\rangle
- \frac{\mathrm{i}}{2} 
\big\langle \Hat{\Bar{\psi}} \star \gamma^\mu \partial_\alpha
\hat{A}_\mu \star \partial_\rho \hat{\psi} \big\rangle
\Big)\bigg)~.
\end{align*}
We must also take the rotational transformation of $\theta^{\mu\nu}$ into
account, 
\begin{align}
W^T_{\theta;\tau}\theta^{\mu\nu} &:= 0 ~,
&
W^R_{\theta;\alpha\beta}\theta^{\mu\nu} &:= 
\delta^\mu_\alpha \theta_\beta^{~\nu} 
- \delta^\mu_\beta \theta_\alpha^{~\nu} 
+ \delta^\nu_\alpha \theta^\mu_{~\beta} 
- \delta^\nu_\beta \theta^\mu_{~\alpha} ~,
&
W^D_\theta\theta^{\mu\nu} &:=  - 2 \theta^{\mu\nu}~,
\end{align}
which acts according to 
\begin{align}
W^?_{\theta}(U \star V) 
=&\left(W^?_{\theta} U \right) \star V 
+ U \star \left(W^?_{\theta} V\right) 
+
 \frac{\mathrm{i}}{2}\left(W^?_{\theta}\theta^{\mu\nu}\right)
(\partial_\mu U) \star (\partial_\nu V)\label{true} \\
W^?_{\theta} \in & \left\{W^T_{\theta;\tau},W^R_{\theta;\alpha\beta}, W^D_\theta\right\}\nonumber
\end{align}
on the $\star$-product in the gluon-fermion vertex 
$\int d^4x\,\langle \Hat{\Bar{\psi}} \star \gamma^\mu\hat{A}_\mu \star
\hat{\psi} \rangle$. This yields the invariance of the noncommutative
Dirac action under complete rotational transformations,
\begin{align}
(W^R_{\hat{A};\alpha\beta} + W^R_{\hat{\psi};\alpha\beta}
+ W^R_{\theta;\alpha\beta}) \hat{\Sigma}_D &=0~,
\end{align}
under the assumption
\begin{align}
W^?_{\theta}\left\{\hat{A}_\mu,\hat{\psi},\Hat{\Bar{\psi}}
\right\}=0~. \label{Null}
\end{align}
The complete dilatational transformation of the action
(\ref{NCD}) is given by
\begin{align}
(W^D_{\hat{A}} + W^D_{\hat{\psi}} + W^D_\theta) \hat{\Sigma}_D 
&=  \int d^4x \,
 m \Big\langle \Hat{\Bar{\psi}} \star \hat{\psi} \Big\rangle ~.
\end{align}
In summary we have 
\begin{align}
W^T_{\hat{A}+\hat{\psi}+\theta;\tau } \hat{\Sigma}_D &= 0~, &
W^R_{\hat{A}+\hat{\psi}+\theta;\alpha\beta } \hat{\Sigma}_D &= 0~, &
W^D_{\hat{A}+\hat{\psi}+\theta} \hat{\Sigma}_D &= -m \frac{\partial
  \hat{\Sigma}_D}{\partial m}~, 
\end{align}
where
\begin{align}
W^T_{\hat{A}+\hat{\psi}+\theta;\tau } &= 
W^T_{\hat{A};\tau } + W^T_{\hat{\psi};\tau } + W^T_{\theta;\tau } ~, &
W^R_{\hat{A}+\hat{\psi}+\theta;\alpha\beta } &= 
W^R_{\hat{A};\alpha\beta } + W^R_{\hat{\psi};\alpha\beta } 
+ W^R_{\theta;\alpha\beta } ~, 
\nonumber
\\
W^D_{\hat{A}+\hat{\psi}+\theta } &= 
W^D_{\hat{A} } + W^D_{\hat{\psi} } + W^D_{\theta } ~.
\end{align}

The Casimir operators (mass and spin) related to the
representation (\ref{WHTF}), (\ref{WHRF}) are 
\begin{align}
m^2 \hat{\psi} &= - g^{\tau\sigma} W^T_{\hat{\psi};\tau} 
W^T_{\hat{\psi};\sigma} \hat{\psi}~, &
s(s+1) m^2 \hat{\psi} &= - g_{\tau\sigma} W^{PL;\tau}_{\hat{\psi}} 
W^{PL;\sigma}_{\hat{\psi}} \hat{\psi}~,
\end{align}
where 
\begin{align}
W^{PL;\sigma}_{\hat{\psi}} \hat{\psi} &= -\frac{1}{2}
\epsilon^{\sigma\tau\alpha\beta} W^T_{\hat{\psi};\tau} 
W^R_{\hat{\psi};\alpha\beta}
= -\frac{1}{8}
\epsilon^{\sigma\tau\alpha\beta} 
[\gamma_\alpha,\gamma_\beta] \partial_\tau \psi
\end{align}
is the Pauli-Ljubanski vector. This yields
\begin{align}
m^2 \hat{\psi} &= - \partial^\tau \partial_\tau \hat{\psi} ~,
\\
s(s+1)m^2\hat{\psi} &= \frac{1}{32} (
g^{\tau\sigma} g^{\alpha\gamma} g^{\beta\delta} 
+2 g^{\tau\gamma} g^{\alpha\delta} g^{\beta\sigma})  
[\gamma_\alpha, \gamma_\beta][\gamma_\gamma, \gamma_\delta]
\partial_\sigma \partial_\tau \hat{\psi}
= - \frac{3}{4} \partial^\tau \partial_\tau \hat{\psi}~,
\end{align}
showing that (\ref{WHTF}), (\ref{WHRF}) is a spin-$\frac{1}{2}$
representation.

\subsection{Seiberg-Witten differential equations}

As in the bosonic case we derive the Seiberg-Witten differential
equations via a covariant splitting of
$W^?_{\hat{A}+\hat{\psi}+\theta}$
\begin{align}
W^?_{\hat{A}+\hat{\psi}+\theta} \equiv 
W^?_{\hat{A}+\hat{\psi}} + W^?_{\theta} &= \tilde{W}^?_{\hat{A}+\hat{\psi}} +
\tilde{W}^?_{\theta} ~,
\label{split}
\\{}
[\tilde{W}^?_{\hat{A}+\hat{\psi}},
W^G_{\hat{A}+\hat{\psi};\hat{\lambda}}] &=
W^G_{\hat{A}+\hat{\psi};\hat{\lambda}^?_{\hat{A}+\hat{\psi}}} ~,
\qquad\qquad
[\tilde{W}^?_{\theta},W^G_{\hat{A};\hat{\lambda}}] =
W^G_{\hat{A};\hat{\lambda}^?_\theta} ~,
\label{covariance}
\end{align} 
whereby we require (\ref{covariance}) to be valid on both
$\hat{A}_\mu$ \emph{and} $\hat{\psi}$.  To find the sought for
splitting we first apply the ansatz of \cite{Bichl:2001yf}:
\begin{align}
\tilde{W}^T_{\hat{A};\tau} &= W^G_{\hat{A}+\hat{\psi};\hat{\lambda}^T_\tau} 
+ \int d^4x\,\mathrm{tr}\Big(
\hat{F}_{\tau\mu} \frac{\delta}{\delta \hat{A}_\mu} \Big) ~,
\label{TTA} 
\\*
\tilde{W}^R_{\hat{A};\alpha\beta} &= 
W^G_{\hat{A}+\hat{\psi};\hat{\lambda}^R_{\alpha\beta}} 
+\int d^4x\,\mathrm{tr}\Big(\Big(
\frac{1}{2} \{\hat{X}_\alpha,\hat{F}_{\beta\mu}\}_\star 
-\frac{1}{2} \{\hat{X}_\beta,\hat{F}_{\alpha\mu}\}_\star 
- W^R_{\theta;\alpha\beta} (\theta^{\rho\sigma})
\hat{\Omega}_{\rho\sigma\mu} \Big)
\frac{\delta}{\delta \hat{A}_\mu} \Big)~,
\label{TRA}
\\*
\tilde{W}^D_{\hat{A}} &= W^G_{\hat{A}+\hat{\psi};\hat{\lambda}^D} 
+ \int d^4x\,\mathrm{tr}\Big(\Big(
\frac{1}{2} \{\hat{X}^\delta,\hat{F}_{\delta\mu}\}_\star 
- W^D_{\theta} (\theta^{\rho\sigma})
\hat{\Omega}_{\rho\sigma\mu} \Big)
\frac{\delta}{\delta \hat{A}_\mu} \Big)~,
\label{TDA}
\end{align}
where we replaced $W^G_{\hat{A};\hat{\lambda}^?}$ by
$W^G_{\hat{A}+\hat{\psi};\hat{\lambda}^?}$.
$\hat{\Omega}_{\rho\sigma\mu}$ is a polynomial in the covariant
quantities $\theta,\hat{F},\hat{D}\dots\hat{D} \hat{F}$, antisymmetric
in $\rho,\sigma$, of power-counting dimension $3$, and expresses the
freedom in the splitting. In the following we set
$\hat{\Omega}_{\rho\sigma\mu}=0$. The parameters $\hat{\lambda}^?$ are
unchanged and given by \cite{Bichl:2001yf} 
\begin{align}
\hat{\lambda}^T_\tau &= \hat{A}_\tau ~,
&
\hat{\lambda}^R_{\alpha\beta} &= 
\frac{1}{4}\{2 x_\alpha+\theta_\alpha^{~\rho}\hat{A}_\rho,
\hat{A}_\beta\}_\star
-\frac{1}{4}\{2x_\beta+\theta_\beta^{~\rho}\hat{A}_\rho,
\hat{A}_\alpha\}_\star~, 
&
\hat{\lambda}^D &=\frac{1}{2}\{x^\delta,\hat{A}_\delta\}_\star~.
\label{lambda}
\end{align}
We write down a covariantization of (\ref{WHTF})--(\ref{WHDF}),
\begin{align}
\tilde{W}^T_{\hat{\psi};\tau} &= \int d^4x \,  \bigg( 
\Big\langle \frac{\overleftarrow{\delta}}{\delta \hat{\psi}} \,
\hat{\mathrm{D}}_\tau \hat{\psi} \Big\rangle
+ \Big\langle 
\hat{\bar{\mathrm{D}}}_\tau \hat{\bar{\psi}} \,\frac{\overrightarrow{\delta}}{
\delta \hat{\bar{\psi}}} \Big\rangle \bigg) ~, 
\label{WHTFT}
\\
\tilde{W}^R_{\hat{\psi};\alpha\beta} &= \int d^4x \,  \bigg( 
\Big\langle \frac{\overleftarrow{\delta}}{\delta \hat{\psi}} \Big(
\frac{1}{4} [\gamma_\alpha,\gamma_\beta] \hat{\psi} 
+ \hat{X}_\alpha \star \hat{\mathrm{D}}_\beta \hat{\psi} 
- \frac{\mathrm{i}}{2} \theta_\alpha^{~\rho} 
\hat{\mathrm{D}}_\rho \hat{\mathrm{D}}_\beta \hat{\psi}
- \hat{X}_\beta \star \hat{\mathrm{D}}_\alpha \hat{\psi} 
+ \frac{\mathrm{i}}{2} \theta_\beta^{~\rho} 
\hat{\mathrm{D}}_\rho \hat{\mathrm{D}}_\alpha \hat{\psi}
\nonumber
\Big) \Big\rangle
\nonumber
\\
& 
+ \Big\langle 
\Big( 
- \frac{1}{4} \hat{\bar{\psi}} [\gamma_\alpha,\gamma_\beta] 
+ \hat{\bar{\mathrm{D}}}_\beta \hat{\bar{\psi}} \star \hat{X}_\alpha 
+ \frac{\mathrm{i}}{2} \theta_\alpha^{~\rho} 
\hat{\bar{\mathrm{D}}}_\rho \hat{\bar{\mathrm{D}}}_\beta
\hat{\bar{\psi}} 
- \hat{\bar{\mathrm{D}}}_\alpha \hat{\bar{\psi}} \star \hat{X}_\beta
- \frac{\mathrm{i}}{2} \theta_\beta^{~\rho} 
\hat{\bar{\mathrm{D}}}_\rho \hat{\bar{\mathrm{D}}}_\alpha
\hat{\bar{\psi}} \Big)
\frac{\overrightarrow{\delta}}{\delta \hat{\bar{\psi}}}
\Big\rangle
\bigg)\,, 
\label{WHRFT}
\\
\tilde{W}^D_{\hat{\psi}}  &=  \int d^4x \, \bigg( 
\Big\langle 
\frac{\overleftarrow{\delta}}{\delta \hat{\psi}} \Big(
\frac{3}{2} \hat{\psi} + \hat{X}^\delta \star 
\hat{\mathrm{D}}_\delta \hat{\psi} 
+ \frac{1}{4} \theta^{\rho\sigma} \hat{F}_{\rho\sigma} \star \hat{\psi}
\Big) \Big\rangle
\nonumber
\\
&\hspace*{4em}
+ \Big\langle
\Big(\frac{3}{2} \hat{\bar{\psi}} + 
\hat{\bar{\mathrm{D}}}_\delta \hat{\bar{\psi}} \star x^\delta 
- \frac{1}{4} \theta^{\rho\sigma} \hat{\bar{\psi}} \star 
\hat{F}_{\rho\sigma}  \Big)
\frac{\overrightarrow{\delta}}{\delta \hat{\bar{\psi}}}
\Big\rangle
\bigg)~.
\label{WHDFT}
\end{align}
where the covariant coordinates \cite{Madore:2000en} are defined by
$\hat{X}^\mu=x^\mu+\theta^{\mu\nu}\hat{A}_\nu$. We define
$W^?_{\hat{A}+\hat{\psi}}$ as the sum of (\ref{TTA})--(\ref{TDA}) with
(\ref{WHTFT})--(\ref{WHDFT}).  Now it is easy to evaluate
$\tilde{W}^?_\theta=W^?_{\hat{A}+\hat{\psi}+\theta}
-\tilde{W}^?_{\hat{A}+\hat{\psi}} =
W^?_\theta(\theta^{\rho\sigma})\frac{d}{d \theta^{\rho\sigma}}$, with
\begin{align}
\frac{d}{d \theta^{\rho\sigma}} 
= \frac{\partial}{\partial \theta^{\rho\sigma}} 
+ \int d^4x\,\bigg(\mathrm{tr} \Big( \frac{d \hat{A}_\mu}{
d \theta^{\rho\sigma}} 
\frac{\delta}{\delta \hat{A}_\mu} \Big)
+ \Big\langle \frac{\overleftarrow{\delta}}{\delta \hat{\psi}} 
\frac{d \hat{\psi}}{d \theta^{\rho\sigma}} 
\Big\rangle
+ \Big\langle 
\frac{d \hat{\bar{\psi}}}{d \theta^{\rho\sigma}} 
\frac{\overrightarrow{\delta}}{\delta \hat{\bar{\psi}}} 
\Big\rangle \bigg)~,
\end{align}
which yields the Seiberg-Witten differential equations
\begin{align}
\label{SWA}
\frac{d \hat{A}_\mu}{d \theta^{\rho\sigma}}& =  
-\frac{1}{8} \big\{ \hat{A}_\rho, \partial_\sigma \hat{A}_\mu +
\hat{F}_{\sigma\mu} \big\}_\star 
+\frac{1}{8} \big\{ \hat{A}_\sigma, \partial_\rho \hat{A}_\mu +
\hat{F}_{\rho\mu} \big\}_\star~,
\\[1ex]
\frac{d \hat{\psi}}{d \theta^{\rho\sigma}}
&= - \frac{1}{4} \hat{A}_\rho \star \partial_\sigma \hat{\psi} 
+ \frac{1}{4} \hat{A}_\sigma \star \partial_\rho \hat{\psi} 
+ \frac{\mathrm{i}}{8} [\hat{A}_\rho,\hat{A}_\sigma]_\star \star
\hat{\psi} 
\nonumber
\\
& \equiv -\frac{1}{8} \hat{A}_\rho \star
\big(\partial_\sigma \hat{\psi} + \hat{\mathrm{D}}_\sigma \hat {\psi} \big)
+ \frac{1}{8} \hat{A}_\sigma \star
\big(\partial_\rho \hat{\psi} + \hat{\mathrm{D}}_\rho \hat {\psi} \big)~,
\label{SWF}
\\
\frac{d \hat{\bar{\psi}}}{d \theta^{\rho\sigma}}
&= - \frac{1}{4}  \partial_\sigma \hat{\bar{\psi}} \star \hat{A}_\rho 
+ \frac{1}{4} \partial_\rho \hat{\bar{\psi}} \star \hat{A}_\sigma 
+ \frac{\mathrm{i}}{8} \hat{\bar{\psi}} \star 
[\hat{A}_\rho,\hat{A}_\sigma]_\star 
\nonumber
\\
& \equiv -\frac{1}{8} \big(\partial_\sigma \hat{\bar{\psi}} 
+ \hat{\bar{\mathrm{D}}}_\sigma \hat {\psi} \big) \star \hat{A}_\rho 
+ \frac{1}{8} \big(\partial_\rho \hat{\bar{\psi}} 
+ \hat{\bar{\mathrm{D}}}_\rho \hat {\psi} \big) \star \hat{A}_\sigma~.
\end{align}
The differential equation (\ref{SWA}) was first found in
\cite{Seiberg:1999vs}. The equation (\ref{SWF}) was for noncommutative
QED to lowest order in $\theta$ first obtained in \cite{Bichl:2001gu}.
It follows from the algebra given in \cite{Bichl:2001yf} (extended to
include fermions) that $\tilde{W}^?_\theta$ satisfies automatically
the second identity in (\ref{covariance}), $[\tilde{W}^?_\theta,
W^G_{\hat{A}+\hat{\psi};\hat{\lambda}}] =
W^G_{\hat{A}+\hat{\psi};\hat{\lambda}^?_\theta}$ so that the
$\theta$-expansion of the action (\ref{NCD}) is invariant under
commutative gauge transformations. One checks the identity
\begin{align}
\Big[W^?_{\hat{A}+\hat{\psi}+\theta},\theta^{\rho\sigma}
\frac{d}{d \theta^{\rho\sigma}}\Big]=0  
\end{align}
for the theory enlarged by fermions, which means that the
$\theta$-expansion based on (\ref{SWA}) and (\ref{SWF}) leads to a
commutative action invariant under commutative rotations and
translations and with commutative dilational symmetry broken by the
mass term.

\section{Conclusion} \label{con}

Following the ideas of \cite{Bichl:2001yf} we have constructed a
representation of the infinitesimal rigid conformal transformations
for noncommutative fermion fields. We have shown that the requirement
that the individual operators of $\theta$- and
$(\hat{A},\hat{\psi})$-transformations commute with gauge
transformations up to another gauge transformation leads directly to
the $\theta$-dependency of the fermion fields first found in
\cite{Bichl:2001gu}.

\end{document}